# Submission to the Commission on Taxation and Welfare on Introducing a Site Value Tax

Eóin Flaherty[1]

October 2021

## Executive summary

The Commission's mandate includes examining the merits of a site value tax, an item that has been present on the Irish public policy agenda since at least 2004. This submission advocates the introduction of a site value tax in Ireland. A site value tax is synonymous with a land value tax. It is a tax on the unimproved value of land, excluding the value of buildings and other infrastructure on it. As such, it is a tax on wealth. The most practical way to introduce a site value tax would be to change the basis of the local property tax and commercial property rates systems from whole-property values to site values.

Site values are easier for valuers to estimate than whole-property values. The work of the Valuation Office could be re-directed to create rolling estimates of commercial and residential site values at the same time. The government could also consider transferring the responsibility for collecting commercial rates from local authorities to the Revenue Commissioners as part of this reform. This would be in the spirit of the current civil service objective of creating efficiencies through shared service arrangements.

The current policy framework is not leading to affordable outcomes in the property market. House prices and rents have been characterised by price growth and volatility that have far exceeded those in employee wages, construction costs and consumer price inflation in the last 25 years. The evidence suggests that these features also apply to the commercial property market.

Extensive property price growth and volatility constrain broader social and economic development. While government fiscal policy has been used to make housing more affordable this has been through increased government expenditure rather than fiscal policy to reduce property price growth. Since government revenue has not grown as fast as property price growth, reducing property price growth is a more sustainable method of ensuring property affordability.

One of the legacies of the 2007 Financial Crisis has been a widespread consensus on the importance of macroprudential economic policy. Site value taxation is an excellent macroprudential fiscal policy. It has the potential to reduce property price growth and volatility and be a sustainable revenue source throughout the economic cycle. The Central Bank's mortgage credit limits are representative of macroprudential financial policy and have been effective in dampening house price growth. However, more can be done through macroprudential fiscal policy.

---

[1] Central Statistics Office (CSO) and University College Dublin. The views and suggestions expressed here are my personal responsibility and not those of my employer. I thank David Flaherty, Constantin Gurdgiev, Ronan Lyons and James Pike for very helpful comments on this paper. I also thank Constantin Gurdgiev, Ronan Lyons, Emer Ó Siochrú and James Pike for their statements of support of it. These statements are included at the end of the document.

Ireland's price level is the second highest in the EU. This means that while the average Irish person's current expenditure is higher than the EU average, their real (price-adjusted) expenditure is below it. High property prices are an important cause of this. Reducing property price growth makes property more affordable to individuals and firms. Lower commercial property prices would increase Irish firms' cost competitiveness. There has been relatively little lending to Irish-owned firms since the Financial Crisis. Relative decline in property prices relative to income would also result in other forms of productive investment becoming more attractive such as investing in small and medium enterprises.

Site value tax has many other benefits. It captures price gains due to the community and government rather than owners' efforts and thus diminishes the incentive to buy land for speculative reasons. Site value tax encourages sustainable land use by encouraging the development of the most attractive sites, as to leave them unoccupied would be costly. Since land cannot be moved to a tax haven, it is also a difficult tax to avoid or evade. Site value tax can be used to finance major infrastructural investments, help facilitate site assembly for development and as a support for the maintenance of protected structures. Site value tax is also a tax on wealth.

A site value tax is a progressive tax since those holding the most valuable sites are eligible for the highest levels of taxation. This is particularly the case from a commercial perspective where sites most conducive to largest commercial gains would be eligible for the highest tax payments.

However, while the evidence suggests that the prices of private dwellings are strongly correlated with incomes, it would not necessarily be always progressive on an income basis. There are several potential ways that could be considered to ensure ability to pay based on current income. These include exemptions or tax credits for those lowest paid, allowing homeowners to defer payments until after the sale of their property, and exceptions for properties with the lowest site values.

The Commission has an important role in helping to develop public discourse on this topic and should use this opportunity to propose the introduction of site value taxation.

# Introduction

The Commission on Taxation and Welfare has called for submissions on tax reform in Ireland. The Commission's mandate includes examining the merits of a site value tax to achieve housing policy objectives in the context of policy sustainability, the experience of previous interventions and the current State supports for housing provision.[2] A site value tax is synonymous with a land value tax. It is a tax on the unimproved value of land, excluding the value of buildings and other infrastructure on it.

This submission advocates the introduction of a site value tax in Ireland. The most practical way to do so would be change the basis of the local property tax and commercial property rates systems from whole-property values to site values. The State could also consider transferring the responsibility for collecting commercial rates from local authorities to the Revenue Commissioners as part of this reform. This would be in the spirit of the current civil service objective of creating efficiencies through shared service arrangements.

Currently commercial rates are valued based on rolling estimates by the Valuation Office while local property tax is based on self-reported values for 1 May 2013 for homes existing at that time. Site values are much easier for valuers to estimate than whole-property values. The work of the Valuation Office could be re-directed to create rolling estimates of commercial and residential sites at the same time.

Site value tax has been present on the Irish public policy agenda since at least 2004 (NESC, 2004). It formed part of the 2009 Programme for Government, the 2010 EU-IMF Programme of Financial Support for Ireland and the 2011 Programme for Government. It has frequently been proposed by the National Economic and Social Council (NESC) since 2015.[3] There have also been many papers and reports supporting site value tax in Ireland since the mid-2000s. These include Pike (2006), Gurdgiev (2009a), Gurdgiev (2009b), Monaghan (2010), Lyons (2011), Collins and Larraghy (2011), Ó Siochrú (2012), An Taisce (2012), Urban Forum (2013), National Competitiveness and Productivity Council (2015), Society of Chartered Surveyors Ireland (2018), IBEC and Property Industry Ireland (2018) and OECD (2018). These papers and reports have tended to focus on site value tax on zoned residential and commercial land.

Discussion of site value tax has a longer history in Ireland. It was advocated by Michael Davitt in the context of the reform of Irish land ownership from the late 19th century onwards (Ó Siochrú, 2011). Its merits were also discussed by the Kenny Report (1973).[4]

---

[2] The full text is that it will "consider the appropriate role for the taxation and welfare system, to include an examination of the merits of a Site Value Tax, in achieving housing policy objectives. This consideration should include reviewing the sustainability of such a role. It should also have regard to the experience of previous interventions in the housing and construction market and the current significant State supports for housing provision." See: https://www.gov.ie/en/organisation-information/7cf49-commission-on-taxation-and-welfare-2021-terms-of-reference/
[3] See NESC (2015), NESC (2018) NESC (2020) and NESC (2021).
[4] Further details on the Kenny Report (Report of the Committee on the Price of Building Land) can be found in the appendix.

## Property prices in broader context

### House prices

Housing in Ireland has become very expensive, both for rental accommodation and house prices (Figure 1).

House rental and house prices have far outstripped growth in wages and consumer prices. This has made housing much less affordable. The cost of renting has increased annually by 4 percent, resulting in an overall growth of 147 percent since 1996. Furthermore, house prices increased by almost twice as fast as incomes in the last 25 years. House prices have increased on average by 10.2 percent per year, resulting in an overall increase of 254 percent since 1996. When we compare ourselves to 1996, wages may have grown by 129 percent, but rents have grown by 147 percent and house prices have grown by 254 percent.

A second aspect of the price of housing in Ireland is its volatility. In 2007 wage growth had doubled relative to 1996 while house prices had quadrupled. However, 2012 house prices had halved in value compared to the 2007 peak, returning to the trend of wage growth. Since then, house prices have again increased by 80 percent while wages have only grown by 18 percent.

While it has not been as stark as the pattern for house price growth, the cost of renting has also been volatile. The cost of renting grew at a slower rate than wages up to 2007, increasing by 86 percent, compared to the wage increase of 102 percent. It fell by 17 percent between 2007 and 2012. The cost of renting has increased by 60 percent since 2012 while wage growth has only been 18 percent.

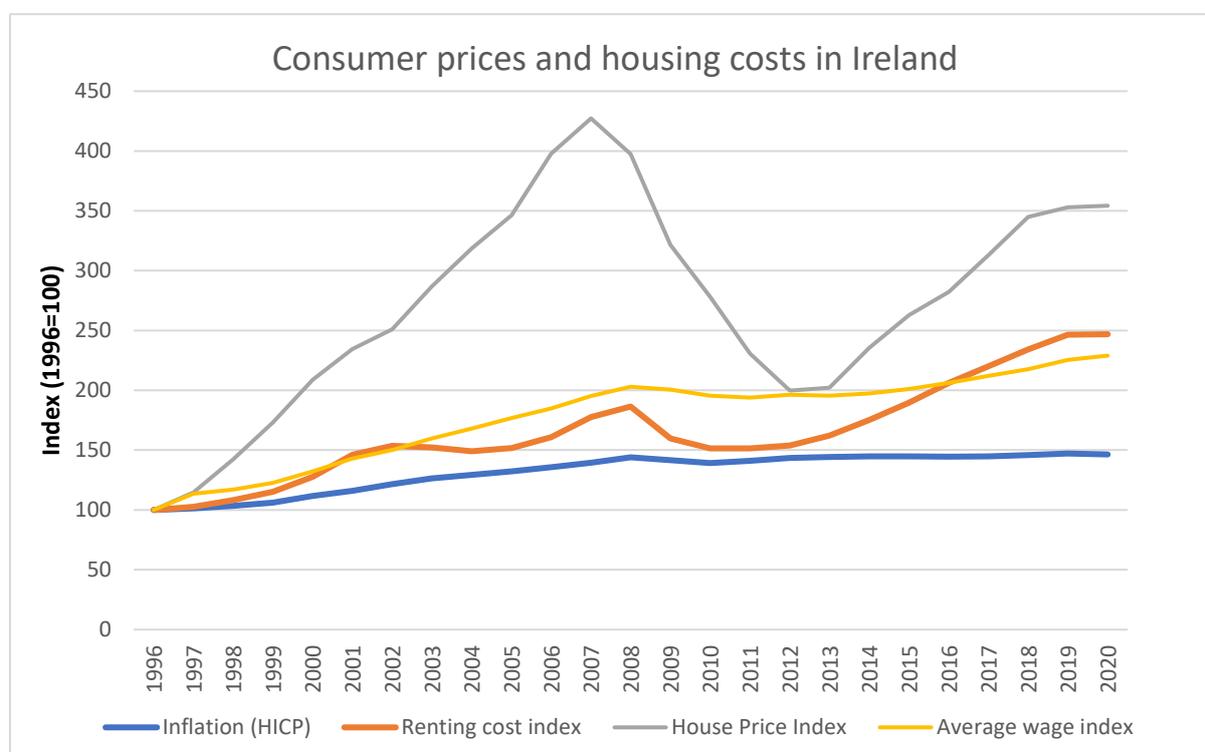

*Figure 1 Source: Eurostat and CSO*

Consequently, average house prices have varied dramatically relative to household wages. In 2006 the average house price was eight times household wages. In 2012 it was 4.6 times household wages while in 2019 it increased to six times household wages (see Figure 11 in appendix).

Both the increases and volatility in house prices is almost entirely determined by changes in site values. Housing can be divided into two components; site value and the capital asset (i.e. the built structure) on the site. Figure 2 displays this data for the average residence in Ireland. The average house price in Ireland was approximately 367 thousand euros in 2007. In 2012, this value fell to 184 thousand euros while in 2019 it increased to 325 thousand. Most of the price change is associated is driven by changes in site values. In 2006, the site value of the average house was 204 thousand euros while the capital asset was worth 163 thousand euros. In 2012, the average site value was worth only 81 thousand euros, a decline of 60 percent of its 2006 value. The value of the capital asset had declined by 36 percent to 103 thousand euros; a difference that was mostly driven by both falling replacement costs and depreciation. In 2019, the average house increased to 325 thousand euros. The value of the capital asset increased to 132 thousand euros while the land value increased to 193 euros. This follows what we should expect. In the long run, land value appreciates while capital assets depreciate.

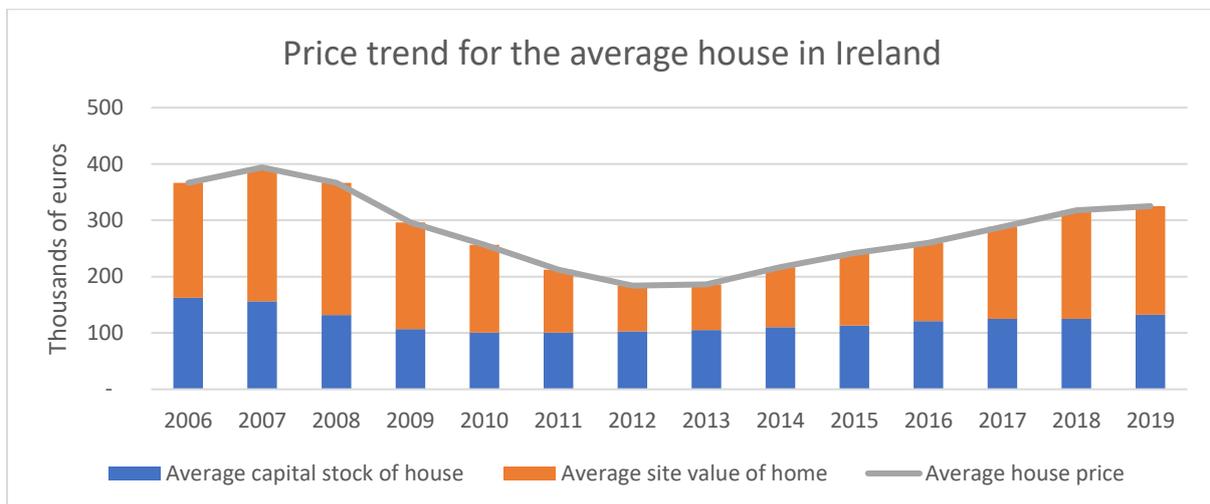

*Figure 2 Source: CSO and author's calculations.*

The importance of site values can be seen more starkly when we look at the contributions to house price growth every year. Figure 3 shows that, excepting 2008 and 2020, changes in site values were by far the largest contributor to changes in house prices.

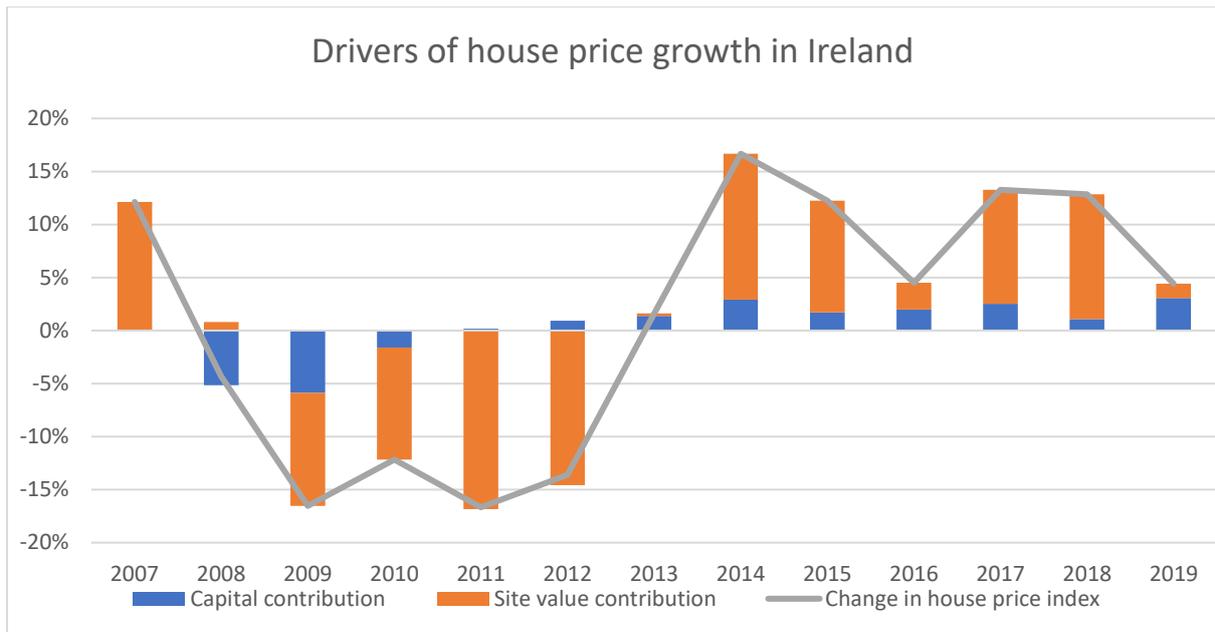

*Figure 3 Source: CSO and author's calculations.*

## Commercial property prices

These issues are also relevant for commercial property. However, the equivalent data is not currently available.[5] The data that is available suggests similarly stark trends for commercial property. By 2011, rent on both industrial and prime office space in Dublin fell to approximately half their Celtic Tiger peak. By 2018, prime office rent exceeded its 2007 peak levels while industrial rent had recovered strongly although not to its 2008 peak.

---

[5] Unfortunately we are still missing publicly available official statistics on commercial property both in Ireland and internationally. Further information on the situation in Ireland can be found on the Central Statistics Office website: https://www.cso.ie/en/methods/methodologicalresearch/commercialproperty/
Further information available on the international situation can be found on the Bank for International Settlements website: https://www.bis.org/ifc/publ/ifc_report_cppis.pdf

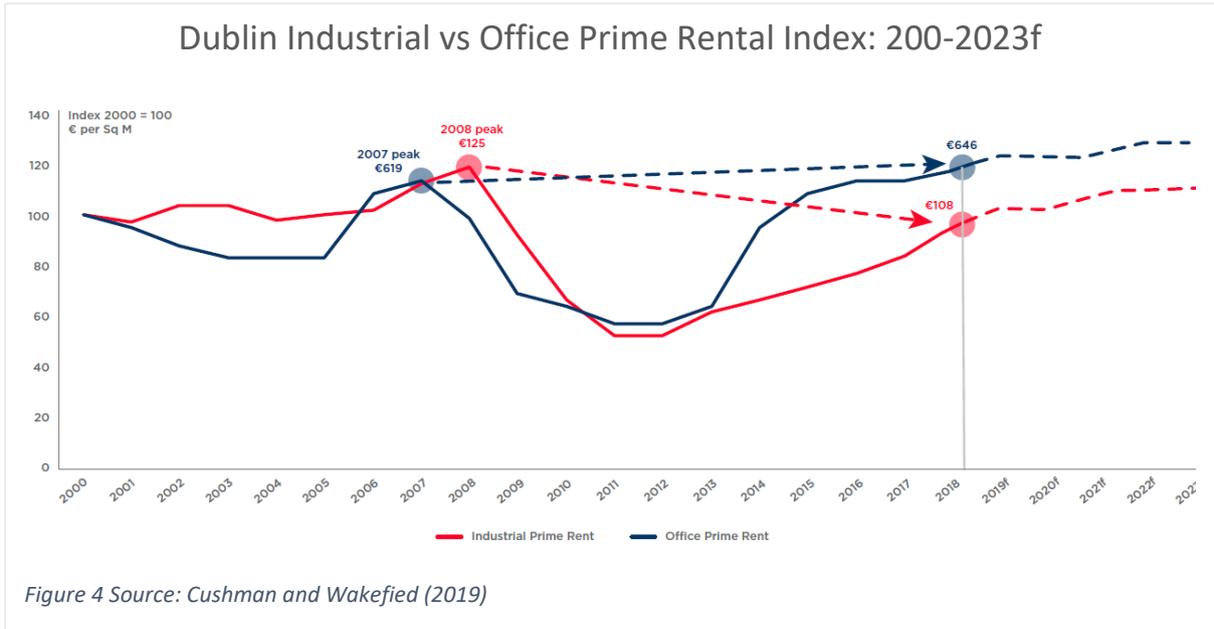

*Figure 4 Source: Cushman and Wakefied (2019)*

Similarly, the National Competitiveness and Productivity Council noted in 2020 that Dublin's rental price of prime office space was the second most expensive compared to a selection of European cities, ranking only below London and Paris. It was ranked above cities such as Milan, Amsterdam, Madrid, Berlin and Vienna (National Competitiveness and Productivity Council, 2020).

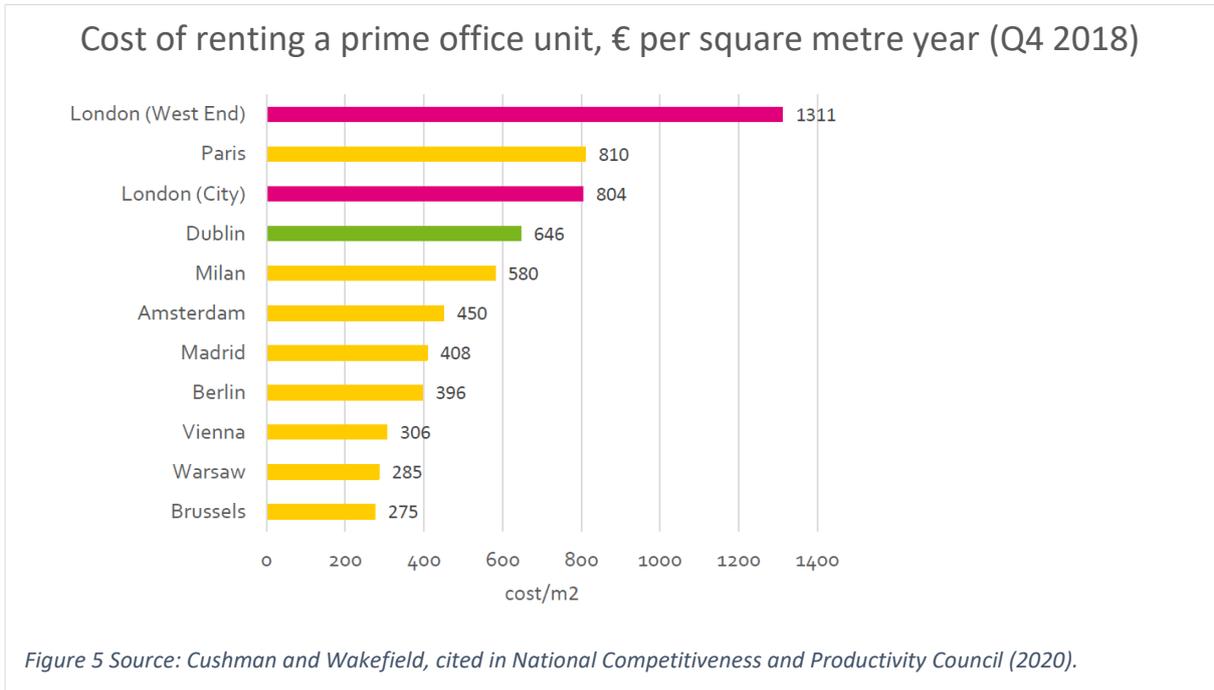

*Figure 5 Source: Cushman and Wakefield, cited in National Competitiveness and Productivity Council (2020).*

## Social Consequences

The rising gap between house prices and wages is the largest contributor to the dynamic behind Thomas Piketty's observation that the wealth to income ratio is rising rapidly in capitalist countries back to levels last seen in the 19th century (Ryan-Collins, 2018). This observation is also made by Joseph Stiglitz who advocates taxing site values to reduce the inequality caused by this phenomenon (Stiglitz, 2015a and 2015b). Ryan-Collins (2019) points out that high and rapidly rising site values makes life chances increasingly determined by whether one is lucky enough to own a piece of land in the right part of the country rather than hard work, innovation or entrepreneurial endeavour.

Corrigan et al. (2019) provide evidence on housing affordability for those who are renting or have a mortgage in Ireland in 2016. They find that average affordability for such households was slightly higher in 2016 than 2006, almost the peak of the last property bubble.[6]

Using CSO Survey on Income and Living Conditions (SILC) microdata, they find that on average, households were paying 20 percent of their income on housing costs in 2016, a slight increase compared to the pre-boom peak in 2006. However, certain groups faced significant affordability challenges. In particular, one in six (16 percent) households had high or unaffordable housing costs.[7] This was even higher for subgroups of households. For example, one in three (32 percent) renters had high or unaffordable housing costs and one in four households in the bottom 20 percent of income earners had high or unaffordable housing costs.

The 25 percent of households on the affordability distribution were paying over 40 percent of their income on housing. Those who paid most of their income were private sector renters, people living in Dublin and mid-east regions, and low income households. Households with high mortgage costs had higher rates of mortgage arrears and of consistent poverty, and lower levels of residual income. Complementing these findings, recent CSO data suggests that almost one in two (47%) of those who live alone in rented accommodation said that they are often under financial pressure.[8]

Negative though some aspects of this picture may be, it belies the broader social consequences of increasingly expensive housing relative to incomes. Since it is an analysis of housing affordability for households currently renting or with mortgages, Corrigan et al. (2019) does not give an indication of some of the broader consequences of the considerable increase in the cost of housing relative to incomes.

Households are defined as a group of individuals living in a housing unit. Except for those in arrears, by definition, households renting or with a mortgage have succeeded in finding housing that the household is able to pay for. For example, the analysis in Corrigan et al. (2019) does not take account individuals or families for whom the cost of housing is too high, forcing them to share a household with others or live in locations that are increasingly inconvenient for them to live.

One consequence of expensive housing is that people are priced out of the accommodation they would otherwise live in. This evidenced by the fact that, in Ireland, younger adults are living with their parents for longer than before. In 2019, 78 percent of young adults aged 16-29 lived with their parents, up from 71 percent in 2007 (See Figure A6, appendix). Ireland had the seventh highest share of 16-29 year olds living with their parents in the EU in 2019 (see Figure A7, appendix).

These high levels of young adults living at home has increased the apparent affordability of housing as measured by Corrigan et al. (2019). Where parents are renting or still paying off a mortgage, young adults' presence in their parents' homes boost the aggregate income of the household. This feature would reverse if they moved out, as well as push up the cost of housing. Recent data indicates that nearly nine in ten (88%) respondents living with a parent said they would prefer to move out, while one in two (50%) parents living with an adult child would like the adult child(ren) to move out.[9]

---

[6] A summary of previous studies can be found in Corrigan et al. (2019).
[7] They define housing costs as high-cost or unaffordable if they exceed 30 percent of income. This is an internationally-used definition.
[8] https://www.cso.ie/en/releasesandpublications/fp/fp-pslhrlpla/pulsesurvey-lifeathome2021rentersloneparentsandadultslivingaloneorwithaparent/

Another trend that belies growing housing costs relative to incomes since 1996 is increased household income associated with the considerable increase in female participation in the labour market over the same period. The share of women in employment in the economy increased by a fifth between 1996 and 2019.[10] Similarly, the gender pay gap and the gender gap in working hours has declined over the period.[11] These factors have likely increased both the average number of household incomes and the average amount of income that households earn.

A further factor that is not captured in affordability data alone is the extent of government expenditure on housing and housing supports. Government expenditure on housing was 2.4 billion in 2019. The largest category was 1.1 billion on social housing (mostly local authority housing, followed by 165 million on homeless accommodation). Separately, the government spent 656 million euros on supporting renters in private accommodation, meaning that eight in every 100 euros spent on private rent in 2019 was paid by central government.[12] Of the remainder, 299 million was spent on house purchase assistance while 233 million was spent on construction grants.

## Economic Consequences

Ireland is currently the second most expensive country in the EU (Figure 7). This means that while the average Irish person's current expenditure is 9th in the EU and higher than the EU average (Figure 19, appendix), their real (price-adjusted) expenditure is the 12th in terms of relative consumption (Figure 6) and below the EU average.[13] Costs are the main reason for Ireland's relatively low real consumption expenditure. High residential and commercial property prices are undoubtably an important part of this.

---

[9] https://www.cso.ie/en/releasesandpublications/fp/fp-pslhrlpla/pulsesurvey-lifeathome2021renterslonparentsandadultslivingaloneorwithaparent/
[10] It was 47 percent in 2019, up from 39 percent in 1996 (see Figure 16, appendix).
[11] Indices for these items were not readily available for the whole period. However, see the following:
https://www.cso.ie/en/releasesandpublications/ep/p-wamii/womenandmeninireland2019/work/
https://www.cso.ie/en/releasesandpublications/ep/p-hes/hes2015/ebg/
[12] Private expenditure on rent was 7.8 billion in 2019. Government rent subsidies (HAP, RAS and SCHEP) were 656 million in 2019. Sources: Department of Public Expenditure and Reform and CSO.
[13] Ireland's relatively low real consumption per capita may be surprising. It is also made by Honohan (2021) using World Bank data for the year 2018.

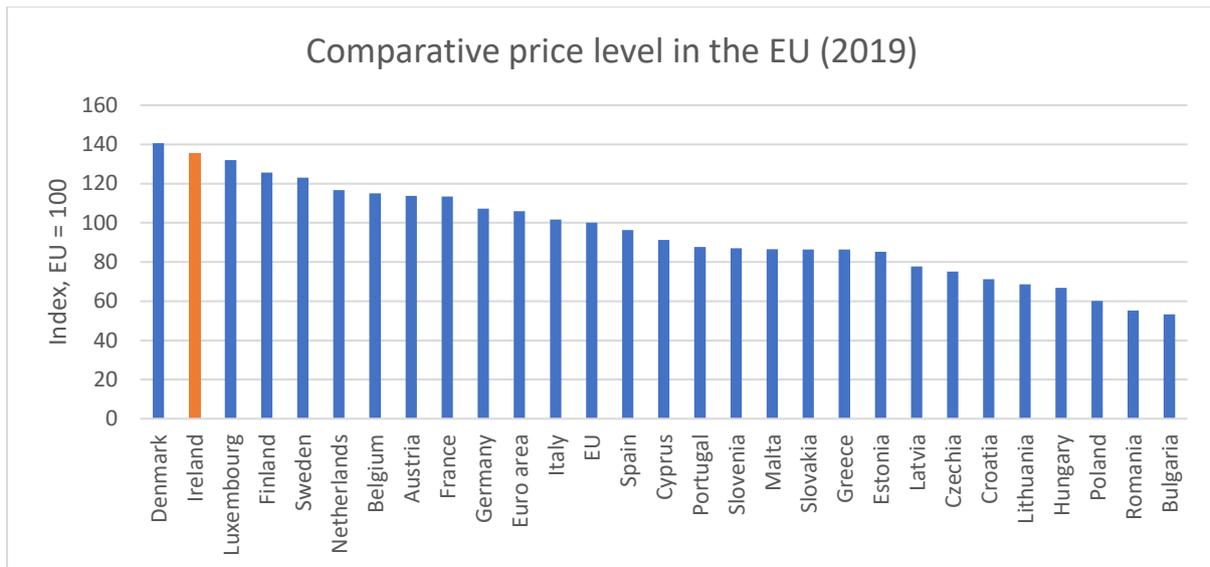

*Figure 6 Comparative price levels are the ratio between Purchasing power parities (PPPs) and market exchange rate for each country. PPPs are currency conversion rates that convert economic indicators expressed in national currencies to a common currency, called Purchasing Power Standard (PPS), which equalises the purchasing power of different national currencies and thus allows meaningful comparison. If the index of the comparative price levels shown for a country is higher than 100, the country concerned is relatively expensive as compared with the EU average. Source: Eurostat.*

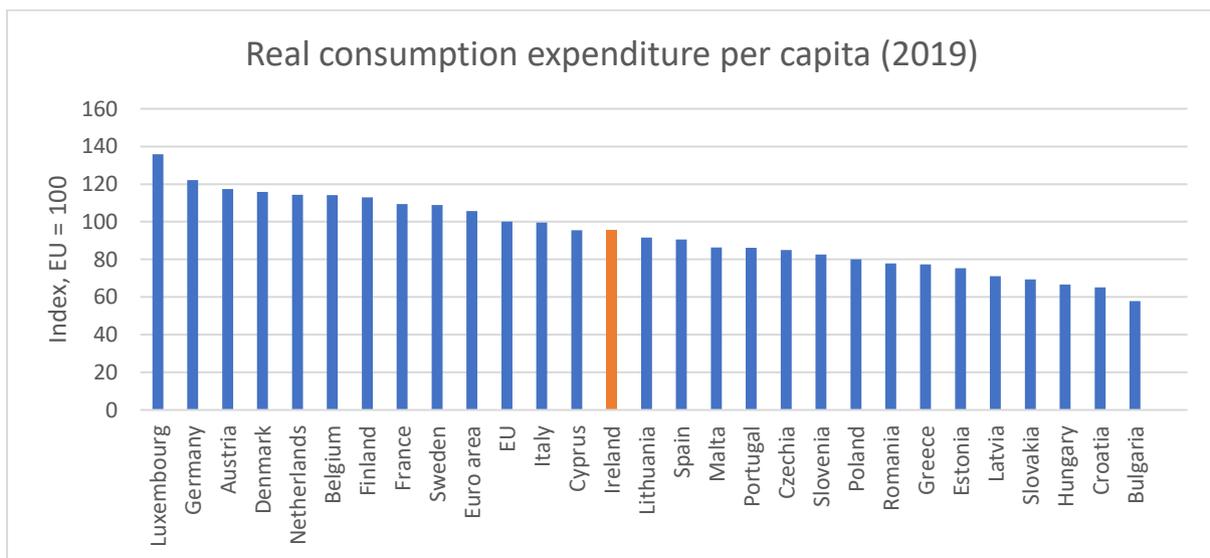

*Figure 7 Actual individual consumption by country per capita. Source: Eurostat.*

High property prices increase costs in other parts of the economy, reducing competitiveness and crowding out other economic activity. Housing is usually by far the largest asset that people purchase in their lives. Rent or mortgage payments are often the most expensive item in a consumer's basket of goods. High housing costs require higher wages for workers, increasing the cost of labour for firms. Higher housing costs also cause increased government expenditure on housing supports for individuals and families.

Higher commercial property prices reduce firms' cost competitiveness in a similar way. Higher prices mean that firms must pay more on rent and on acquiring property. Higher commercial property prices are also a clear barrier to entry for new firms and other firms without commercial property, reducing competition and benefiting the firms that are already most dominant in the market. Finally, higher property prices divert lending and entrepreneurial activity away from productive activity to the financial and real estate sector. These factors are exacerbated by commercial property prices outstripping the pace of economic growth.

## What sustainable property prices might look like

It seems reasonable to suggest that we should aim for economic sustainability, sustainable government tax revenue and increased overall welfare for society. These objectives would be better served by low property prices and property price stability.

There is no definition in common circulation of what low or sustainable property prices should be. Perhaps the Central Bank's macroprudential "loan to income" mortgage lending limit of 3.5 times one's annual income is an appropriate definition of housing affordability.[14] Currently the average house price is approximately 7.7 times the average employee's gross income and six times average household wages.[15]

Price stability means low price growth and is commonly defined as an annual rate of close to but under two percent.[16] However, we have had average annual house price growth of 6.1 percent with a standard error of 2.5 percent over the last 25 years.

## Site value tax as a solution

Property prices in Ireland are characterised by high price growth and considerable price volatility, and are becoming increasingly unaffordable. Site value taxation has the capacity to ensure low property prices and property price stability. This makes it an excellent macroprudential fiscal policy. It has the potential to reduce property price growth and volatility, and to be a sustainable revenue source throughout the economic cycle.

Introducing a site value tax would place downward pressure on the price of land, reducing property price growth.[17] Mirlees et al. (2011) points out that supply is fixed and cannot be affected by the introduction of a tax. The incentive to buy, develop or use land would not change. Site value tax reduces the incentive to buy land for speculative purposes rather than productive purposes since the tax creates a cost to holding land. Property taxes that include buildings are less effective as they are effectively a tax on investment in the property.

In addition to placing downward pressure on the price of property, site value tax increases property price stability (Muellbauer, 2005). This is supported by econometric evidence and applies for other property taxes, but is particularly the case for site value tax (Blöchliger, 2015). The Central Bank's macroprudential limits (limits on mortgage lending relative to the private borrower(s) income and relative to the value of the property) have been effective in dampening house price growth (Acharya et al., 2020). However, since they only affect mortgage provision for private individuals, they cannot be expected to be a panacea and more can be achieved by site value tax as a macroprudential fiscal policy.

---

[14] https://www.centralbank.ie/consumer-hub/explainers/what-are-the-mortgage-measures
[15] These are defined as Compensation of Employees divided by number of employees and households in the country respectively. Comparing median values might be more appropriate than mean values but the data is not readily available to do this.
The median loan-to-income of first-time buyers would have a much lower series, as a result of the Central Bank's macroprudential limits. This indicates that the macroprudential rules work but do not prevent other parts of the income distribution from being priced out.
[16] See, in particular, the mandate of the European Central Bank.
[17] As well as reducing property price growth, site value tax has the potential to even reduce property values. However, given the widespread financial interest in site values staying the same or increasing, this may be too much to expect.

Reducing property price growth makes property more affordable to individuals and firms. The available data suggests that Irish commercial property is expensive relative to other EU countries. Lower commercial property prices would increase Irish firms' cost competitiveness. There has been relatively little lending to Irish firms since the Financial Crisis.[18] Relative decline in property prices relative to income would also result in other forms of productive investment becoming more attractive such as investing in small and medium enterprises (Ryan-Collins, 2018).

Site value tax has many other benefits. It captures price gains due to the community and government rather than owners' efforts and thus diminishes incentive to buy land for speculative reasons. Site value tax encourages sustainable land use by encouraging the development of the most attractive sites as to leave them unoccupied would be costly. It also increases the cost of hoarding land and incentivises it being put to its most valuable use. Taxing sites would reduce the possibility of making capital gains on sites and, if the tax rate was high enough, they would eliminate capital gains entirely. Since land cannot be moved to a tax haven, it is difficult to avoid or evade site value tax (Ryan-Collins, 2018). Site value tax is also a tax on wealth.[19]

A site value tax is a progressive tax since those holding the most valuable sites are eligible for the highest taxation (Lyons and Wightman, 2014). This would particularly be the case from a commercial perspective where sites most conducive to largest commercial gains would be eligible for the highest tax payments.

Site value tax can be used to finance major infrastructural investments, in particular relating to transport. For example, site values in parts of London rose considerably in response to the extension of the London Underground Jubilee line from Green Park to Stratford in the late 1990s. If a site value tax had applied, there would have been no need for the new lines to be paid for from general taxpayers in unaffected areas. Instead, a bond issued for the purpose could have been paid back using the revenues from the uplift in land values.[20]

Site value tax can also facilitate site assembly (while still consistent with discouraging speculation). For example, an industrial estate that society decides would be better used as mixed commercial and residential use could be easily steered to redevelopment through site value tax (by the tax reflecting its new best use). Existing owners would have an incentive to sell. This would avoid the need for complicated or lengthy processes such as compulsory purchase orders.

A site value tax can also be used to support the maintenance of protected structures. Where a protected structure is being maintained, the site value would be close to zero in most cases, as the most valuable use of the site is determined by the older building on it. Where a protected structure is being intentionally allowed to decay in order to free up the site to another use, it could be charged full value (i.e. as if nothing were on the site). This could be done actively. For example, proof of maintenance would need to be shown on an annual basis in order to keep the assessed value low.

---

[18] See Figure 18, appendix. Indigenous Irish firms' debt levels have been declining since 2017. Data for 2020 not yet available. Source: CSO (2020).

[19] Real estate makes up by far the largest fraction of Irish wealth. Site values make up most of the value of residential real estate in Ireland. While the data is not readily available, it is probable that site values also make up most of the value of non-residential real estate. See *Comment by Ronan Lyons* below for further discussion on this topic.

[20] See Wetzel (2012) for discussion and references to studies on this.

Alternatively, it could be done retrospectively. For example, if a property were to fall into a state beyond repair, site value tax could be charged for the preceding 20-year period, plus penalties and charges.

## Implementation

Ireland's current tax system is mostly based on the purchase of products and (employee and corporate) income tax flows.[21] In doing so, taxation increases the price of products, the cost of labour and the cost of declared profits. Taxing site values diversifies the source of government revenue from income flows to a stock of value. Taxing site values widens the tax base and does not discourage economic activity or distort investment decisions.

Ireland has several taxes on property. These currently make up four percent of the overall tax take. Commercial rates are the most important of these at 1.7 percent (1.3 billion). The next is stamp duty at 1.4 percent (1.1 billion). Local property tax is worth 0.5 percent (426 million) while the non-principal private residence charge is worth 0.03 percent (24 million).

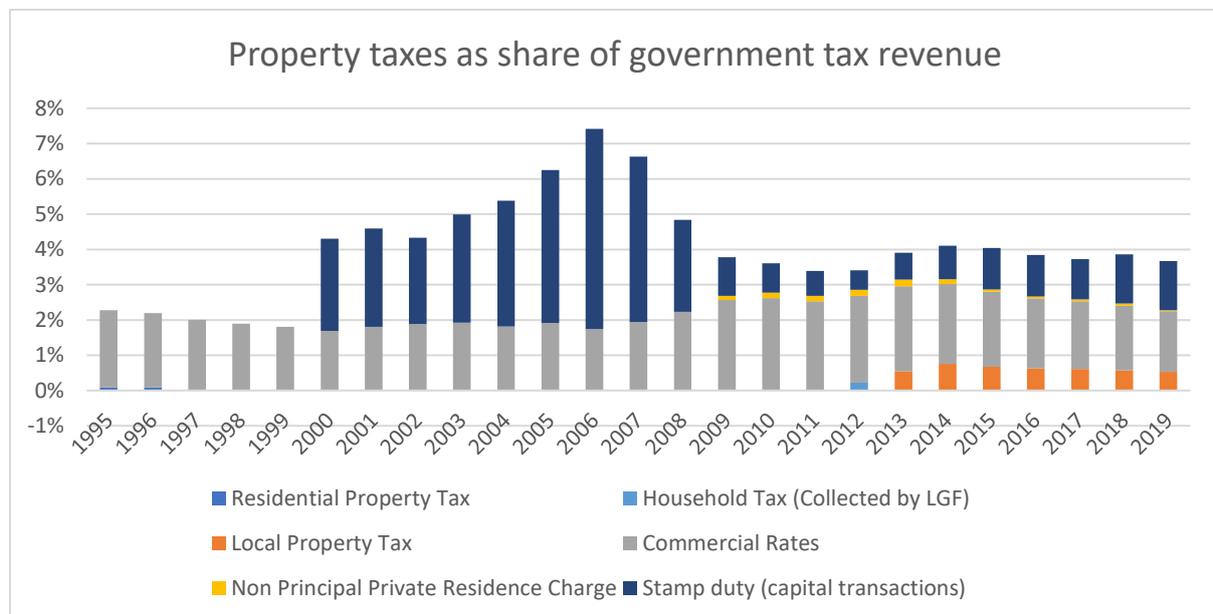

*Figure 8 Source: Eurostat*

---

[21] Just under half (48 percent) of Irish government tax revenue comes from income tax and PRSI. The intuition here is that employees pay according to their means, with the employer paying a share of PRSI. This category is tiered so that one's lower pay is less heavily tax, making this tax progressive, in that higher earners pay progressively more. Higher income tax increases the cost of labour. This increases the cost of production, reduces competitiveness and discourages employment. The next highest source of taxation is VAT, at 22 percent of government tax revenue. This is mostly levied at the same rate of 23 percent, with some exceptions related to goods that the government wants to either encourage consumption of, such as the services of hotels and restaurants. Higher VAT rates increases the cost of products, discouraging consumption of these products relative to others now, relative to spending in the future and relative to spending abroad. Corporation tax accounts for 16 percent of government revenue and is a tax on firms' profits after depreciation.

### Commercial property

Re-basing commercial rates on site values may be particularly straightforward. It is easier to value sites, which covary across broader areas rather than commercial properties. These include both sites and the buildings on them. The latter are often unique and specific to the sector of the occupying firm.

For the most part, commercial firms attempt to generate profits from the assets they have. Site values of commercial property should be strongly correlated with their potential profitability. Furthermore, both the Irish SME Association (ISME) and IBEC have expressed support for basing commercial rates on site values (ISME, 2021) (IBEC and Property Industry Ireland, 2018).

### Private homes

Currently commercial rates are valued based on rolling estimates by the Valuation Office while local property tax is based on self-reported values for 1 May 2013 for homes existing at that time. Derelict homes and homes built since that date are exempt from taxation. Site values are much easier for valuers to estimate than whole-property values. The work of the Valuation Office could be re-directed to create rolling estimates of commercial and residential sites at the same time.

A site value tax would be a progressive tax since those holding the most valuable sites would be eligible for the highest payments. However, while the evidence suggests that the prices of private dwellings are strongly negatively correlated with income deprivation (Lyons and Wightman, 2014), making it a progressive tax overall, it would not necessarily be always income progressive. No tax can mirror the income tax in being income progressive and site value tax is no exception (McCarthy, 2012).[22] There are several potential ways that could be considered to ensure ability to pay. Exemptions could be made for low income homeowners or other means testing (Blöchliger, 2015). Homeowners could be allowed to defer payments until after the sale of their property. Private property with the lowest site values could be made exempt. Changes could also be introduced simultaneously to make other taxes more progressive.

The Commission has a role in bringing expert views towards the centre of the Overton window. Here are some observations in this context. First, the justification for site value tax is much clearer than property tax. Secondly, this is an opportunity to bring the issue of site values of driver of affordability to the fore.

The 2009 Commission on Taxation (Daly et al., 2009) considered that there was a strong economic role for site value tax as the basis of local property taxation. However, it did not recommend its introduction at that time over whole-property taxation for two reasons. One reason was the challenge of communicating its benefits to the public. Hopefully, this document has illustrated that its benefits are straightforward; that site values are the main driver of property price unaffordability and volatility in Ireland, and that taxing site values would help to alleviate these problems.

A second reason was that implementing site value taxation would be more difficult to implement than a property tax on whole-property values. This concern was echoed in the 2015 Review of the Local Property Tax (Thornhill et al., 2015) and the 2019 Review of the Local Property Tax (Hogan et al., 2019). These concerns should be considerably lessened due to the fact that the local property tax is now in place and due to the considerable increase in data collection in this area including by the

---

[22] Further discussion of the income effects of taxing wealth in Ireland can be found in Lawless and Lynch (2016).

Residential Tenancies Board, the Property Services Regulatory Authority (specifically their work on the Residential Property Price Register and the Commercial Lease Register) and the Property Registration Authority (which has digitised the Land Registry). NESC (2021) also argues that these implementation challenges should not be an overriding obstacle.[23] See Lyons (2011) for an overview of how to introduce a site value tax in Ireland. Monaghan (2010) and Collins and Larraghy (2011) also provide analysis on how to do so.

NESC (2021) also points out that if commercial rates and the local property tax were replaced by a site value tax, the work of the Valuation Office would be redirected to valuing site values. Site values are easier to value for valuers than property values as they are more consistent across small areas than overall property values and do not require detailed analysis of different buildings.

### Transfer responsibility for collection of commercial rates to the Revenue Commissioners

The government should transfer the responsibility for collecting commercial rates from local authorities to the Revenue Commissioners as part of this reform. This would be in the spirit of the public service objective of creating efficiencies through shared service arrangements.[24]

## Conclusion

This paper argues that the Commission should support the introduction of taxation on site values. Specifically, it should propose changing the basis of the local property tax and commercial property rates systems from whole-property values to site values.

The current policy framework is not leading to sustainable outcomes in the housing market. House prices and rents have been characterised by price growth and volatility that far exceed other macroeconomic trends. While there is much less data available for the commercial property market, these features are very likely to be similarly applicable to it.

It seems reasonable to state that we should aim for economic sustainability, sustainable government tax revenue and increased overall welfare for society. These objectives would be better served by low property prices and property price stability.

It is evident that extensive price growth and volatility in the property market have some benefits at certain points in time. Buying, selling and entering into rental contracts at the right time provides financial benefits to many individuals. Similarly, a high number of transactions in the housing market when prices are high can provide large tax windfalls to the government.

However, these benefits can be easily reversed and are undoubtably at the cost of overall welfare. They are the expense of others and damage economic development and economic stability. For every seller at the market price peak there must also be a buyer at the peak. Large tax windfalls from a glut of transactions on overpriced homes are a procyclical source of revenue. When prices rise faster on commercial property and commercial rent than general economic growth, non-property related business activity is crowded out through increased costs, costs that may disproportionately affect new firms.

---

[23] NESC points to five public data sources that could be used to provide more accurate valuation of site values on a permanent basis. Not all of these data sources were available in the past.
[24] https://www.gov.ie/en/publication/efd7f-civil-service-renewal-2030/

Government expenditure is best supported by reliable revenue that is not affected by windfalls which encourage increased unplanned expenditure that may be difficult reduce in later years. Rapid property price growth is unlikely to be equitable and is likely to be particularly damaging to those without property assets in society. Ultimately, as employees, shareholders, consumers and taxpayers, we are all affected by these factors.

Site value taxation has the potential to reduce property price growth and volatility and be a sustainable revenue source throughout the economic cycle. Reducing property price growth makes property more affordable to individuals and firms. The available data suggests that Irish commercial property is expensive relative to other EU countries. Lower commercial property prices would increase Irish firms' cost competitiveness. Relative decline in property prices compared to income would also result in other forms of productive investment becoming more attractive such as investing in small and medium enterprises.

Site value tax has many other benefits. It captures price gains due to the community and government rather than owners' efforts and thus diminishes the incentive to buy land for speculative reasons. Site value tax encourages sustainable land use by encouraging the development of the most attractive sites as to leave them unoccupied would be costly. Since land cannot be moved to a tax haven it is difficult to avoid or evade. Site value tax is also a tax on wealth.

The Commission has an important role in helping to shape public discourse on this topic and should use this opportunity to propose the introduction of site value taxation.

## Note on the data

All data used here is publicly available. All of the single-year comparisons here use 2019 data given the unrepresentative COVID-related circumstances in 2020 and 2021. I have tried to use data from 1996 onwards where possible. The breakdown of house prices by capital and land was estimated by combining several data sources. Further information on the data and calculations are available on request.

Ryan-Collins, Josh. *Why can't you afford a home?*. John Wiley & Sons, 2018.

Society of Chartered Surveyors in Ireland (SCSI). *Rejuvenating Ireland's small town centres - A Call to Action*. 2018.

Stiglitz, Joseph E. "The origins of inequality, and policies to contain it." *National Tax Journal* 68, no. 2 (2015): 425-448.

Stiglitz, Joseph E. "To Fight Inequality, Tax Land." *Bloomberg.com*. 2015. Available at: https://www.bloomberg.com/opinion/articles/2015-03-03/to-fight-inequality-tax-land

Stiglitz, Joseph E. *New theoretical perspectives on the distribution of income and wealth among individuals: Part IV: Land and credit*. No. w21192. National Bureau of Economic Research, 2015.

Urban Forum. *Urban Forum Colloquia Summary Document*. 2014. Available at: https://issuu.com/urbanforum/docs/uf_colloquia_summary_document

Wetzel, Dave in Ó Siochrú, Emer ed. *The Fair Tax.* Shepheard-Walwyn. 2012.

## Comment by Constantin Gurdgiev, Associate Professor of Finance, Monfort College of Business, University of Northern Colorado, USA

Having engaged in research into economics of Land Value and Site Value Taxation policies with specific application to Ireland since 2009, I fully welcome and support an excellent Submission to the Commission on Taxation and Welfare on Introducing a Site Value Tax by Eóin Flaherty.

Since the start of this century, Ireland sustained a series of real estate and real estate finance crises, virtually unparalleled to its prior history.

The first crisis was manifested in rapid price appreciation in domestic real estate markets during the period of 2002-2006 that (1) generated price dynamics in property markets that were completely detached from personal income dynamics, resulting in severe affordability crisis in first-time-buyer segment of the markets; (2) created unsustainable push and pull factors within Irish banking system to provide funding for real estate development, investment, speculation and owner-occupier purchases; (3) produced unsustainable, and historically and peer-comparatively unprecedented scale of risk accumulation by Irish lenders and households; and (4) resulted in high degree of fiscal policy coupling to the fortunes of the highly volatile property markets. That crisis culminated in the economically and socially disastrous implosion of Irish financial system in 2008-2010 and contributed to significant fiscal policy pressures, not limited to, but strongly related to the catastrophic fall-off in property market-related tax revenues.

The second crisis developed over the subsequent period of 2011-2019, when a combination of deleveraging pressures stemming from the fallout and mismanagement of the prior crisis resulted in (1) unsustainable rise in rents, followed and accompanied by (2) rapid rise in house prices, (3) chronic shortage of new development in the housing sector in Ireland, (4) rapidly rising homelessness and lack of affordable housing for younger households, and (5) rising economy-wide tax burden stemming from the need to replace revenues lost to the collapse of the property sector.

The third crisis emerged on foot of the second one when the Covid19 pandemic induced a massive shock to employment and household income prospects of many Irish families, while simultaneously disrupting the housing markets supply and demand patterns.

These events have resulted in a deep and protracted banking sector crisis, an unprecedented collapse in Irish building and construction sector, accompanied by decades of run-away house prices appreciation and rents inflation. The fallout from these three crises reaches beyond grave personal and household implications, altering the fabric of the Irish society as a whole (problems of local tenure and continuity of tenure for younger households, reduced physical and social mobility, increased risk to future incomes and pensions, as well as rising risk of homelessness) and the macroeconomic foundations of the Irish economy (putting at risk future tax revenues and pushing up demand for social housing, social services and public health). The rents and house price inflation pressures have resulted in a systemic misallocation of labour and human capital resources economy-wide, with younger workers unable to efficiently match jobs demand to labour supply due to cost pressures and lack of affordable and efficiently distributed housing. In younger demographic cohorts, lack of affordable housing for students attending Irish third level institutions over time can pose a threat to efficiency with which Irish third level education system can function as an effective platform for development of future human capital. These pressures have significantly reduced and are continuing to reduce Irish economy's competitiveness in its ability to attract organic (as opposed to tax-optimising) foreign direct investment, retain on-shore domestic productive investment, reduced the space for Irish indigenous entrepreneurship, and resulted in a depressed domestic demand.

Mis-allocation of public and private resources associated with dysfunctional property markets and runaway property price inflation and rents inflation also present a significant challenge to Ireland in the context of climate change risks and Government's efforts to mitigate such risks. Market failures in the form of excessive costs associated with renting and owning homes in economically and socially optimal locations lead to inefficient use of transportation and the stock of housing, contributing to higher environmental costs absorbed by the society at large. Excessively high cost of housing also reduces the scope for development and deployment of environmental taxation tools, as it erodes overall income base available for such taxation.

This problem has been highlighted by the post-Covid19 pandemic recovery experience, where labour shortages across key domestic sectors are following geographic patterns of housing affordability distribution, with pools of surplus labour concentrated in smaller towns, while demand for labour is growing in major cities. As the result, today's Ireland is facing a risk of severe skills and human capital supply mismatch with demand for skills and labour emerging across a range of sectors in Ireland as the result of decades-long disruption to the functioning and the efficiency of the housing markets. In addition, Ireland is experiencing a persistent and growing problem of homelessness, in part driven by high costs of rental properties. These, and other adverse shocks to housing in Ireland are occurring at the time when there is a substantial stock of incomplete and/or derelict housing units across the country.

In simple terms, Irish property markets are severely lacking proper incentives for efficient and market demand-responsive use of real estate – from farmland left under-utilized, to development land banks sitting idle, to derelict and empty houses left to decay across major cities, to low levels of utilization of mixed-use properties in town and city centres. This lack of effective and efficient incentives for optimization of the property utilization stems, in part, from the poor design of property taxation, which is neither linked to the actual property values, nor to the economic aspects of property use.

An associated result of the above market failures is the ineffective system of revenue raising operating in the property sector. Current tax system of commercial rates and LPT provides no effective mechanism to incentivize more efficient and productive use of land. The same system penalizes private investment in land improvements and efficiency-enhancing economic investments in buildings and other private amenities. At the same time, neither rates nor LPT offer an opportunity to capture private gains or losses sustained as the result of public investments in shared infrastructure and public amenities. The system of property taxation operating in Ireland today is also ineffective in accounting for positive and negative spillovers from land use by proximate and adjoining owners to property owners. In other words, the existent system offers little to no avenues for accounting for private and public externalities that are prevalent in the property markets.

In contrast, international body of evidence shows that Site Value Tax / Land Value Tax (SVT/LVT) represents a taxation framework that is most effective in:

1. Capturing public externalities (positive and negative) arising from adjoining and proximate land use by public amenities that spill over to private owners of land and properties;
2. Capturing private externalities (positive and negative) arising from proximate land and property investments undertaken by other site and property owners;
3. Creating economic value-aligned incentives for additional investment in properties and properties use by existent land users;
4. Creating powerful incentives to pursue socially, environmentally and economically optimal utilization of land and properties over time by property owners;

5. Improving price/cost incentives for real estate investors to pursue investment projects that target improvements in efficiency in land use;
6. Enhancing stability and reducing risks to property-related tax revenues;
7. Creating potentially powerful pathways for linking environmental and social taxation policies to public and private benefits of such policy interventions, without introducing distortionary effects that traditional property-related tax systems have on private and public investments.

These, and other benefits of the SVT/LVT tax framework for property taxation have been outlined, along with supportive evidence, in my earlier work in Gurdgiev (2012(a)), Gurdgiev (2012(b)) and Gurdgiev (2012(c)).

All existent body of evidence surveyed by me over the years shows that SVT/LVT addresses the need for more efficient value-capture mechanisms in the economy. The concept of value-capture "envisions creation of policy tools to adequately capture the privately accruing changes in the value of sites and/or consumption that arise from public infrastructure investments" (Gurdgiev 2012(b)). Gurdgiev (2012(c)) shows that SVT/LVT represents an optimal tax instrument when compared to property tax and the existent structure of property taxation based on the stamp duty, VAT and development charges from the Exchequer perspective, when revenue stability, predictability and counter-cyclicality are of value.

Gurdgiev (2012(b)) introduced and discussed internationally available policies used for raising revenue to finance public investment and used international experience to highlight their main shortcomings and benefits. The basis for this assessment was the view that public investment creates positive or negative values that are dynamically (over time) captured privately in the market pricing of land sites associated with private properties. These values are not perfectly subject to taxation and have nothing to do with private owners own contributions to the property.

Gurdgiev (2012(b)) then draws an argument that SVT/LVT "represents the optimal policy instrument for raising revenue for public investment when compared to all other alternatives. In qualitative rankings, the final distance between the optimal policy (LVT/SVT) and the runner-up policies (Property Tax and Joint Development/Air Rights) is significantly greater than the distance between the least favored two alternatives (Development Impact Fees and Special Assessments). This shows that the economy would gain much greater efficiency from moving from a Property Tax or a PPP-style system of financing (consistent with Air Rights and Joint Development) to a Land Value Tax system of revenue collection, than it would from any other reform within the confines of the above choices of policy instruments."

## Comment by Ronan Lyons, Associate Professor in Economics, TCD

I wholeheartedly support the submission by Eoin Flaherty above. There are three main components to any tax system: taxes on income, taxes on consumption, and taxes on wealth. Real estate in all its forms comprises by far the largest fraction of Irish wealth, with residential real estate alone worth over €500 billion. But, unlike in the case of income or consumption taxes, Ireland's taxation of real estate is not coherent, with no overall coherence across types of real estate or underlying philosophy, such as using the ensuring the taxation of real estate is equitable, efficient and consistent. Instead, the taxation of real estate in Ireland is characterised by a variety of different systems and measures, including stamp duties, commercial rates, local property taxes and development levies, with many types exempt.

The Commission's report presents a unique opportunity to bring consistency to the taxation of real estate, the bulk of Irish wealth, and to ensure that the taxation of real estate supports the State's policy goals, including having a healthy housing system. The decade since the re-implementation of an annual residential property tax in Ireland has been one of the most challenging periods for the Irish housing system since independence. Market rents have doubled, while the sale price of housing has increased by almost three quarters having been falling rapidly at the start of the decade. Despite all of these issues, there still exists no coherent system-wide view within Irish policy on how the housing system should work in delivering appropriate and affordable accommodation for all those living in Ireland. A key element in delivering such a goal lies in the taxation of real estate, both residential and non-residential.

The implementation of a site value tax, in particular on development land and other land not being used for owner-occupied housing would give policymakers significant power to coordinate market forces to deliver socially desired outcomes. There has been significant talk over the last decade about special taxes on vacant or derelict sites, but these have proven tricky to implement and largely ineffectual. More recent attempts to capture upswings in land value due to rezoning are more promising but, even if implemented in full, it would still mean that existing uses are effectively exempt from any consequences relating to their under-utilization of scarce and valuable land, especially in cities. Moving to a site value tax would avoid issues around the legality of taxation-by-use. Moreover, it could be brought in as a revenue-neutral replacement for a range of other charges, including commercial rates, LPT (where applied to owner-occupied housing), development levies and stamp duties. In doing so, it would help move the taxation of real estate, the major form of wealth in Ireland, from being based on more irregular one-off and up-front charges to more regular on-going payments, something which would increase the predictability of Exchequer revenues. More ambitiously, it could be charged at higher level, closer to full economic rents, and be used to replace taxes with greater social costs, including income tax, which disincentivizes work, and VAT, which is regressive.

In terms of implementation, it would of course be a large change in how the taxation system works. For that reason, I would suggest establishing the twin principles of land value literacy and a commitment to a gradual switchover from existing taxes to a site value tax, most probably after some trials on particular sites or in particular areas. In relation to the trials, these could be managed by the Land Development Agency (LDA), which will have large urban sites under its management in the coming years. On land value literacy, i.e. informing everyone - including policymakers, site-holders and citizens - of what land values are, the past decade has seen huge progress in the methods used to calculate values and especially in the relevant data available to policymakers. When it comes to a site value tax that could promote sustainable development, lower the cost of land and thus of new homes, and tax the leading form of wealth in the country, there are no barriers to its adoption other than political will.

## Comment by Emer O Siochru, BArch FRIAI

Eóin Flaherty has eloquently set out the economic and social case for land site value tax and the political and logistical reasons that led to adoption of other mechanisms that have proved unequal to the systemic imbalances in the Irish property sector. I will limit my comments to the finer issues of the main case that he outlined and recent measures under discussion that SVT will impinge upon. The pandemic's impact on the non-residential sectors offers an opportunity to illustrate the benefits of SVT. Commercial office rents and values have been permanently altered by new demands for more home working. The retail sector has winners and losers depending on adaptability to online trading, the hospitality and entertainment sector depending on their location i.e. inner urban fared worse than leafy suburban. The arts and cultural sector, an important component of hospitality, has fared worst of all.

Shifting commercial rates from the users, mostly tenants, to property owners under SVT will give welcome short term relief to productive businesses until the next rent review. Newly reviewed rents will fit the new market reality better than their pre-pandemic comparables. Because SVT is blind to the vacancy of the site or building, the elimination of commercial rates reliefs becomes moot. The arts and cultural sector could be a major beneficiary, filling spaces under temporary free rent licences until commercial tenants emerge. SVT should be expanded to cover all zoned development land at its highest and best use, at the same rate as if built upon and occupied, recognising that it is 'stock in trade' and commercial in nature. This makes complicated a partial windfall taxes redundant and provides the necessary conditions for speedy Community-led Development or reformed Strategic Development Zone (SDZ).

The introduction of SVT on residential uses can start with all rental property on proposed, newly built, and legacy properties and on the investor with a single rental unit to REITS holding thousands of units. This is consistent with the extension of SVT onto zoned land and recognises investment property's role as a 'commodity'. SVT is immune to tax evasion and rebalances the equation between large foreign investors and local at least to some degree. An allowance recognising the 'right to housing' calculated as the affordable price of a modest urban site will be essential to gain homeowner support for the replacement of LPT by SVT on the principal residence. SVT on non-functional rural homes should at least cover their average servicing costs. SVT on Social and Cost Rental Housing should be assessed, rolled up and paid when/if the property is sold to the tenant or in the open market; ditto Seniors over 65 years should be given an option to roll up the payment of SVT until it is sold or inherited.

The strong grassroots movement for Community-led Development includes the Community Land Trust model (CLT) that has been recognised in the recent Housing Bill. Freehold should be retained by the LDA or CLT and perpetual leases granted with affordable ground rents. Fair solutions can evolve to share a % of such ground rent receipts with local government as their SVT contribution.

Growing local and international investment interest in agricultural farmland and forestry land for non-food and non-timber uses strongly demands that they be subject to SVT before they threaten the public interest. Wind turbines and solar photovoltaic are industrial uses while investments in forestry; both 'continuous cover' and 'conventional plantations' for carbon trading purposes are commercial in nature. This conversation is vital for the social sustainability and economic resilience of Ireland as a nation and should be commenced as soon as possible.

The immediate benefit introducing a simple comprehensive SVT is the rapid reduction of the land value/cost element of all homes under all tenure types, existing and new – ditto but to a lesser extent for commercial and agricultural uses. No other single reform can deliver that boon. Finally, SVT's benefit can be compounded by Government using receipts to reduce regressive taxes i.e. VAT and income taxes and perverse charges i.e. stamp duties and development levies.

# Comment by James Pike, Dip.Arch., FRIAI, RIBA

I strongly support the submission made by Eoin Flaherty.

The Fine Gael/Labour Government in 2011 included SVT on residential land in its Program for Government but then decided to adopt the Property Tax instead because when they asked the Revenue Commissioners to administer the SVT, the Commissioners decided to ask the individual owners to value their own sites, but then considered it was impossible for them to value them, as they would only be able assess the total value of the property.

The Property Tax is completely the wrong tax as it penalises the improvement of properties and the full use of sites. SVT was subsequently promoted by Eoghan Murphy, Minister for Housing in the last Government and it was considered by the Cabinet in the Summer of 2019 but they deferred it as they were concentrating on Brexit.

Site Value Tax would be much more effective in reducing the overall cost of housing than any other method under discussion as it would not only reduce the cost of land, but would also remove the need for LPT, Development Levies, Vacant Site Levies and Stamp Duty.

The best way Site Value Tax should be introduced to tax payers is to start with those sectors who support it, and then mitigating the impact on those who can least afford it. Probably the best way is to start by replacing Commercial Rates at their current level. The majority of those now paying them, as represented by the Chambers of Commerce, IBEC and ISME are largely in favour of it. Use the existing local authority staff who currently assess Rates valuations and collect the tax. They would be assisted by the Land Registry Authority.

When tried and tested the Local Authorities would move to land zoned Residential. This would remove the need for the Local Property Tax but also Vacant Site and Development Levies and Stamp Duty.

When Commercial and Residential are both being assessed the process of revaluation could be conducted on alternate years as in Denmark. The cost of assessing SVT in Denmark is 1.5% of the total tax raised.

The cost of SVT for households could be mitigated by reducing Income Tax or VAT, which would have a major benefit for poorer households. The level of SVT could be raised to a higher level with a proportional reduction in Income Tax or VAT.

SVT could be levied on all land, including Agricultural land, with exceptions for areas zoned for Conservation and Biodiversity. This would incentivise the full use of land as many farms are not fully cultivated at present.

The recent interview published in the Business Post with the Chair of the Commission, Prof. Niamh Moloney, was headlined by the fact that the Commission will looking at a wealth tax to pay for the higher health and welfare costs of an ageing population and the costs of reducing carbon emissions. Site Value Tax would probably be the best and most appropriate form of wealth tax, as land is a key element in the funds owned by the wealthiest and many of those approaching, and past retirement age.

# Appendix
## Supplementary charts

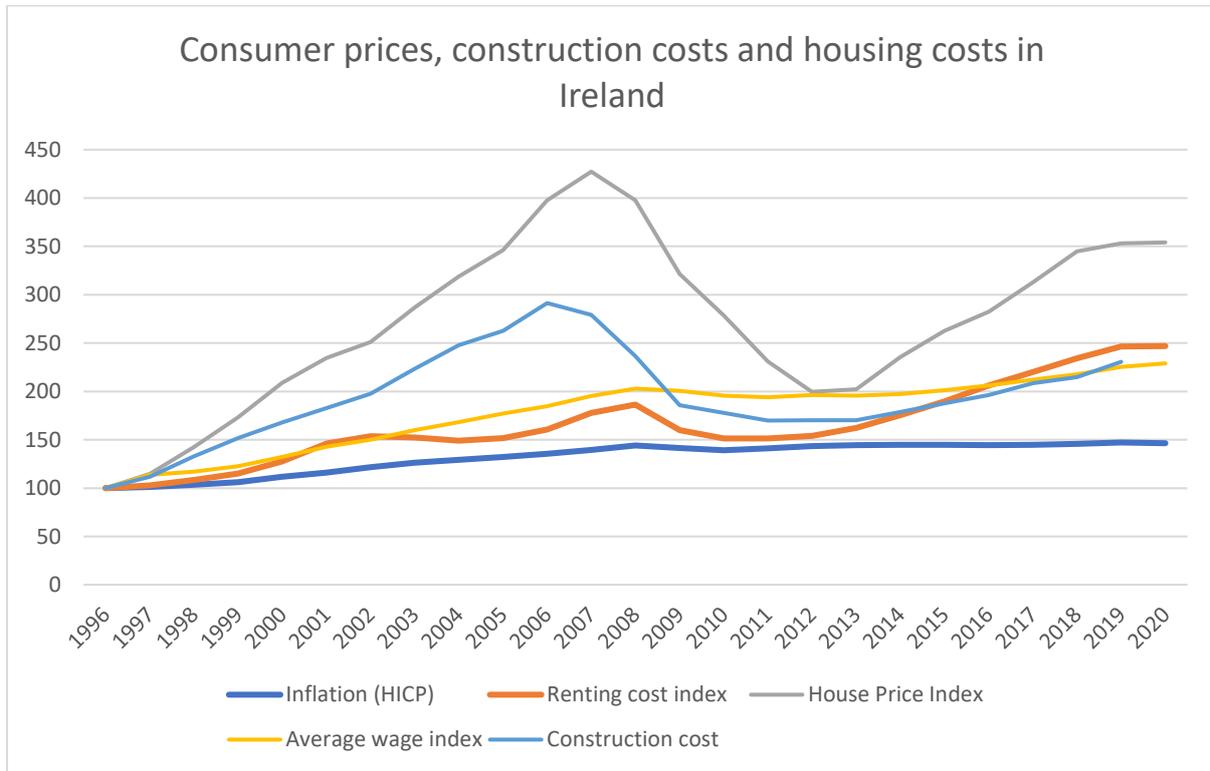

*Figure 9 Source: CSO and Eurostat*

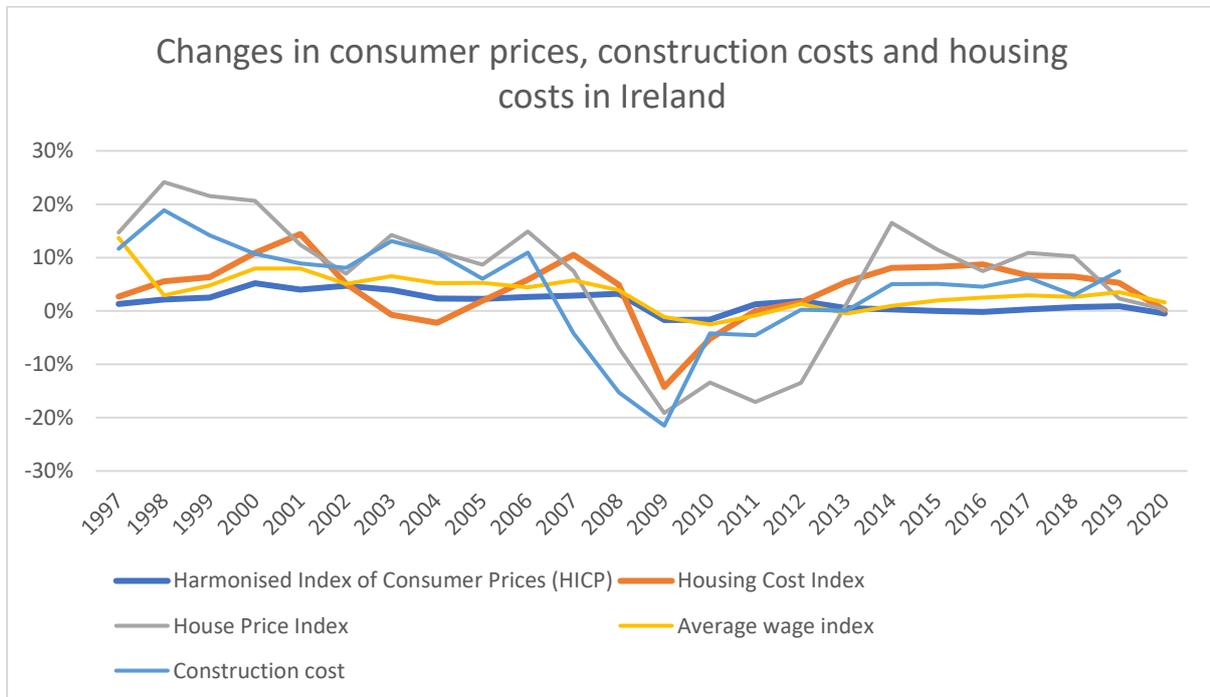

*Figure 10 Source: CSO and Eurostat*

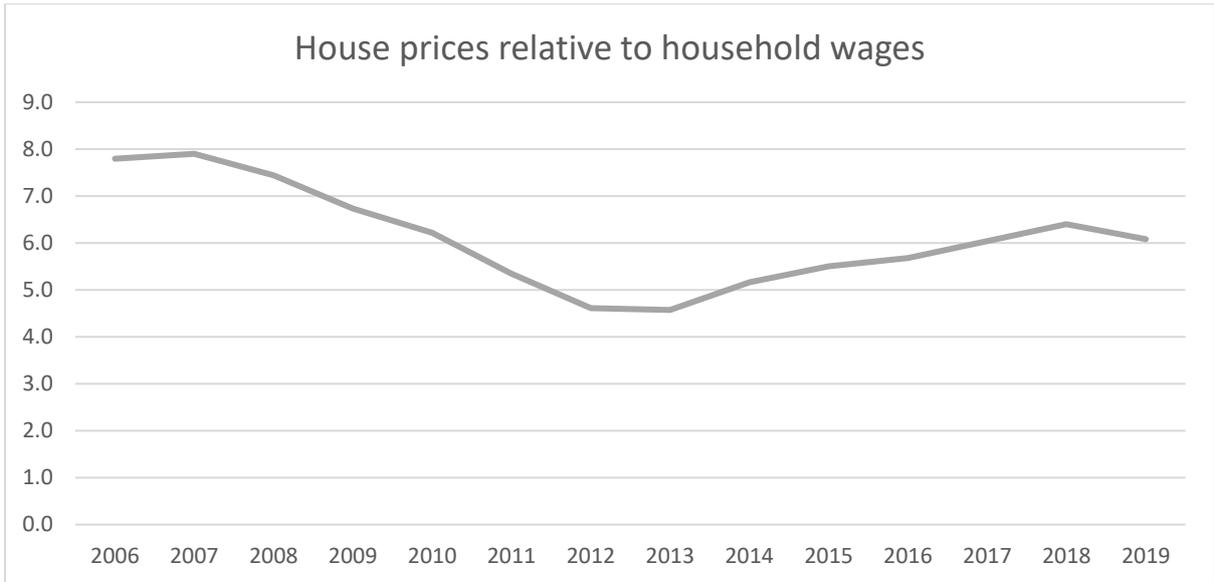

*Figure 11 Sources: CSO, Eurostat and author's calculations*

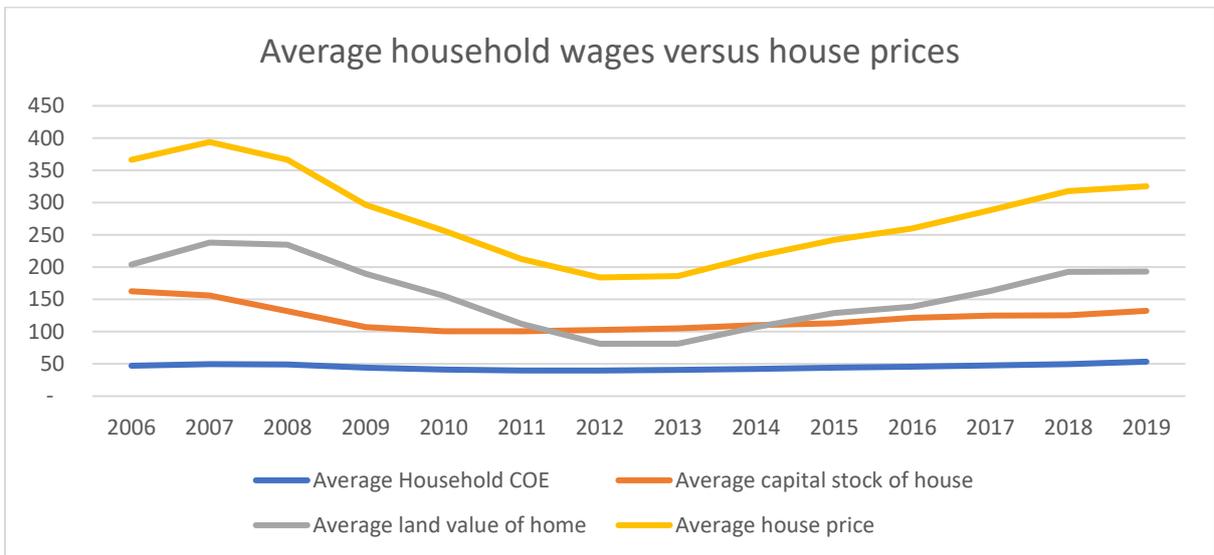

*Figure 12 Sources: CSO, Eurostat and author's calculations*

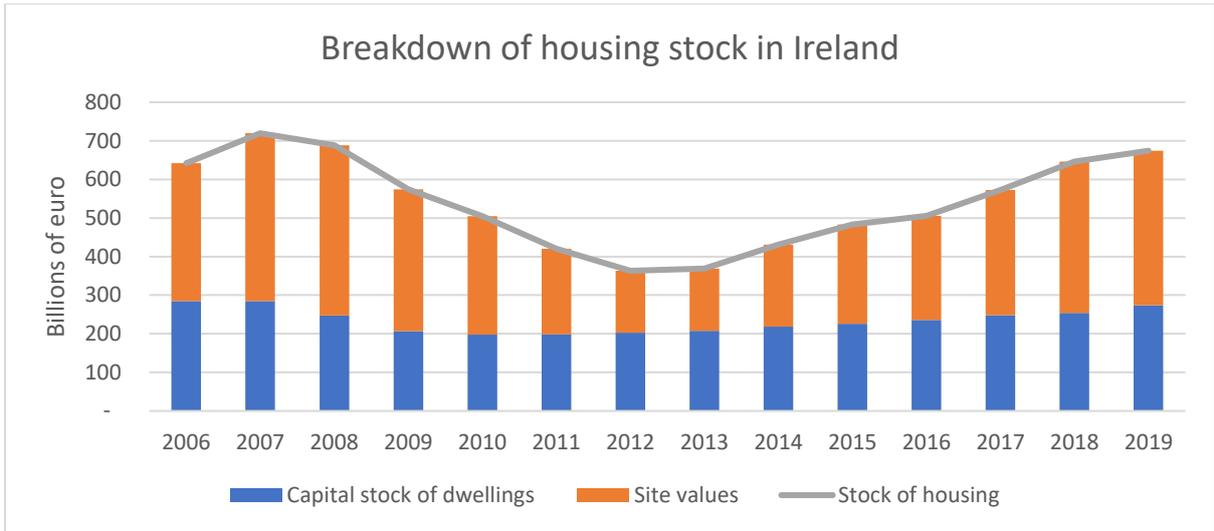

*Figure 13 Sources: CSO and author's calculations*

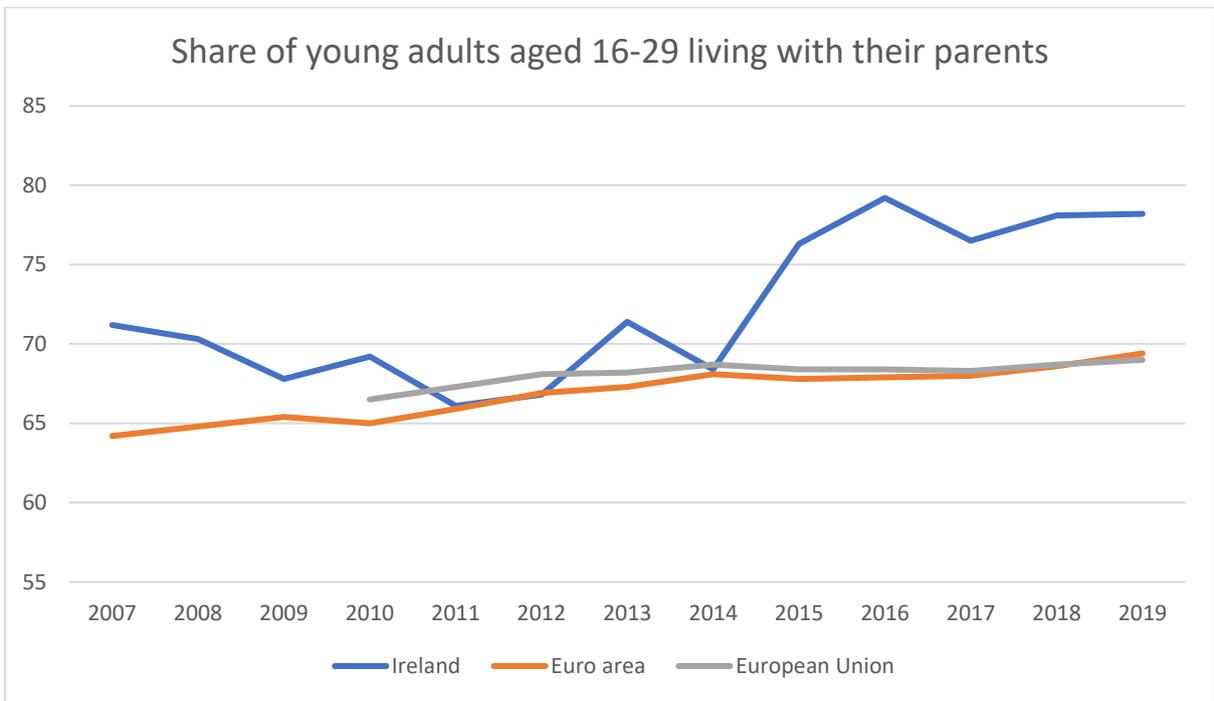

*Figure 14 Source: Eurostat*

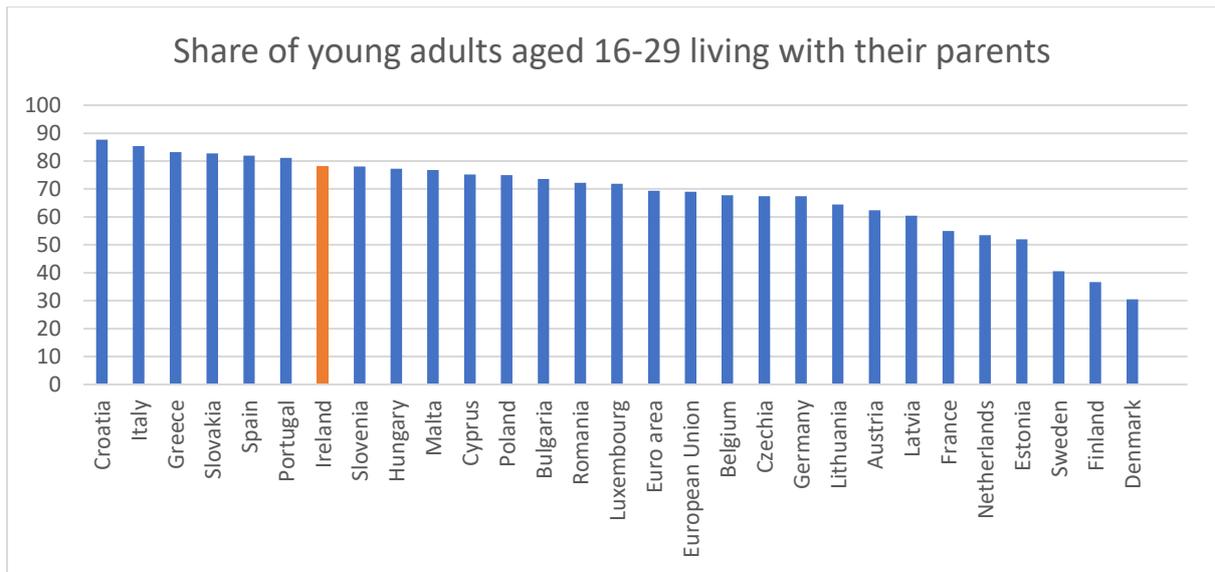

*Figure 15 Source Eurostat:*

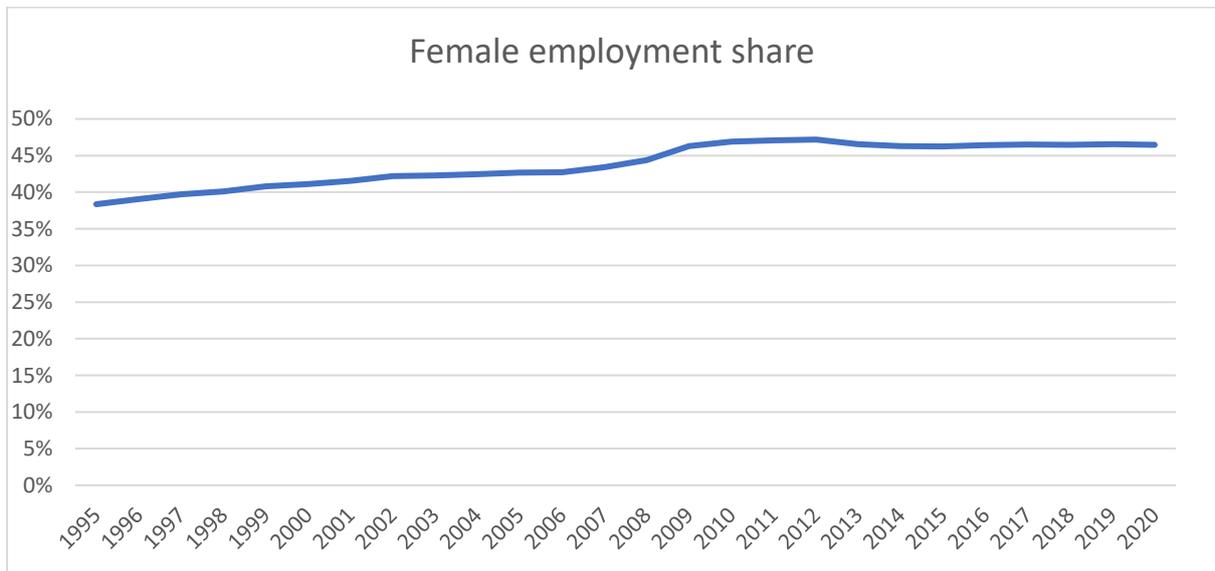

*Figure 16 Source: CSO*

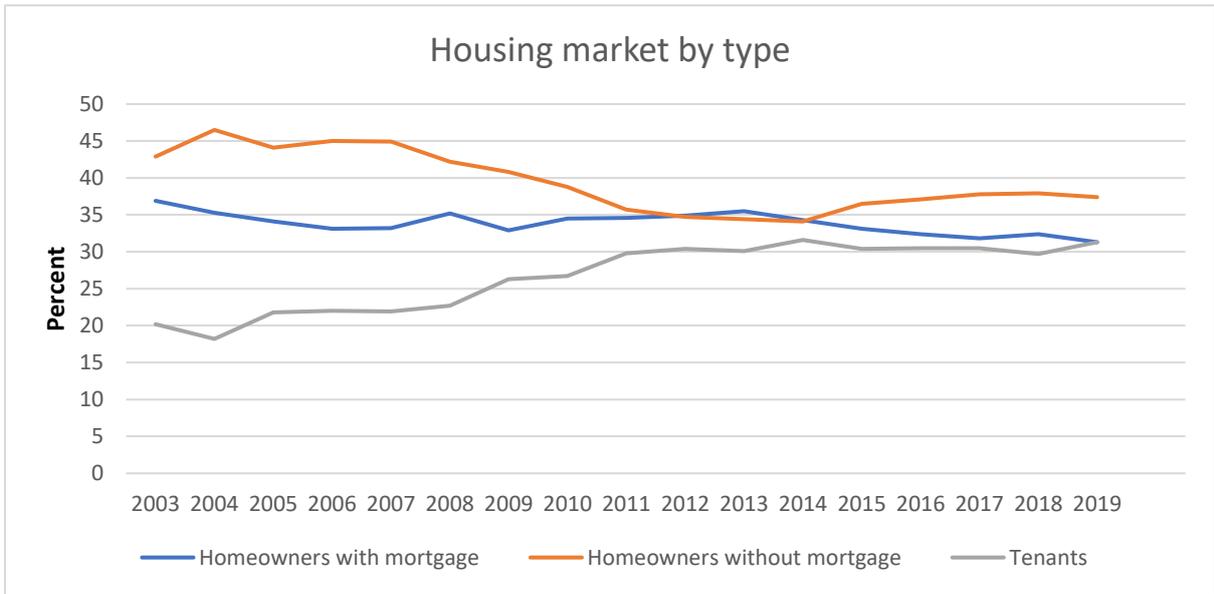

*Figure 17 Source Eurostat*

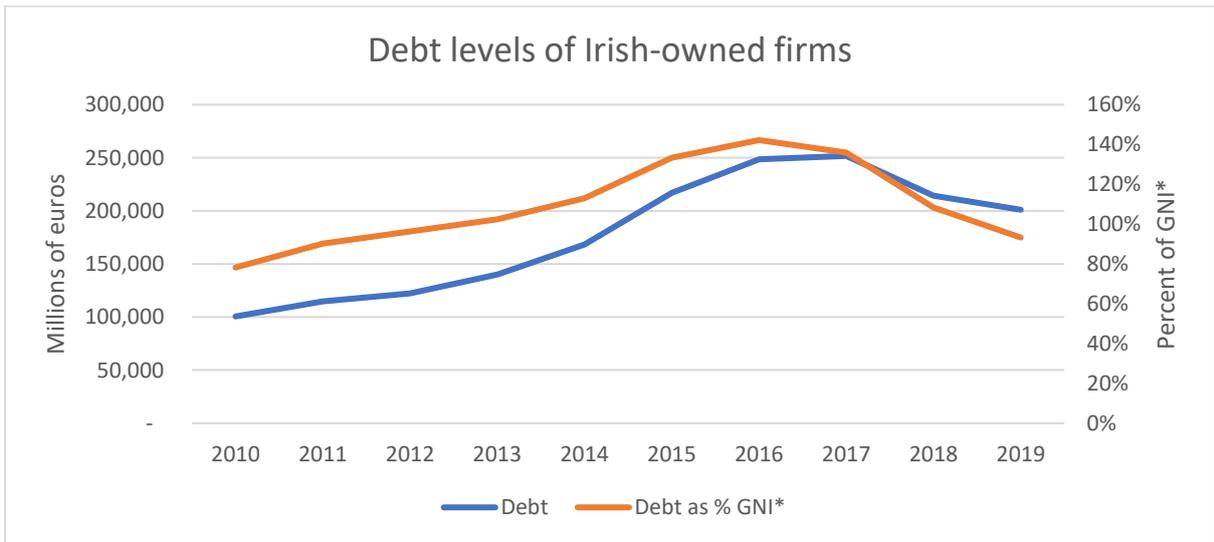

*Figure 18 Source: CSO*

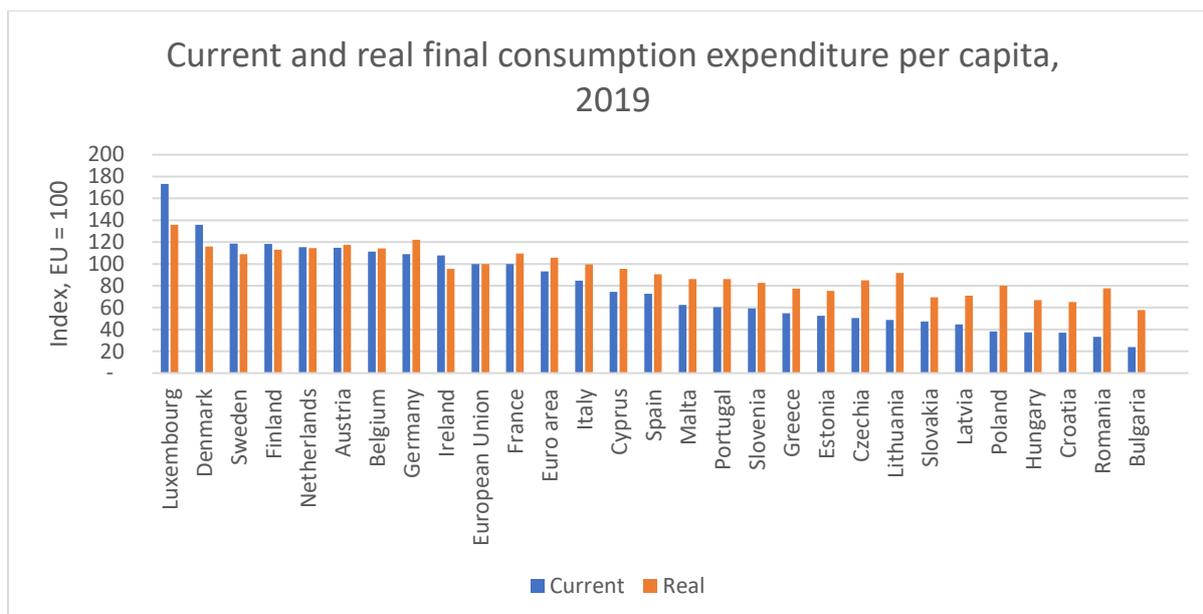

*Figure 19 Source: Eurostat*

### Understanding the economics of land

Land is fixed in supply. This means that it does not follow typical economic theory where increased demand causes increased prices and in turn increased supply. Land is a positional good. This means that the desire for land is related to one's position in society relative to others and not subject to diminishing marginal returns like other goods.

Housing is also different from many other goods. It is a consumption good but is also a financial asset of significant value and is by far the largest asset held by citizens in advanced economies. These features can be complementary under certain circumstances but if house prices (driven by site values) rise beyond incomes, the financial asset function can come to dominate the demand for housing which becomes speculative. In such circumstances, housing becomes less and less available as a consumption good for larger shares of the of the population (Ryan-Collins, 2018).

### House prices and housing supply

It might be assumed that high and rapidly increasing site values are partly a good thing since they incentivise increased supply of housing and commercial property. For example, more expensive housing may encourage more housing construction and house owners to downsize. However, this is not necessarily the case. Quickly rising property prices can potentially even reduce housing supply.

The reason for this is the nature of the sites underlying properties. Site values are the main component of property price change. A site is an asset that is fixed in supply and does not depreciate. Sites also often become more valuable as their location becomes more attractive. Consequently, the basic supply and demand framework does not apply.[25] Increased site values increase the wealth of property owners regardless of whether the property is sold.

---

[25] The basic supply and demand framework can be described using the perfect competition model, usually one of the first things described to economics students. In this model, increased demand causes increased prices, which in turn leads suppliers to increase supply.

Selling, even in a buoyant market, conveys risks. Increased prices may signal further prices increases in the future and encourage postponement of sale. Developing a property entails commercial risks, even in a buoyant market. It also entails the risk of losing out on even higher property prices in the future. A buoyant market does not guarantee that it will be easy to find an appropriate alternative home that suits one's needs.

Quickly rising property prices are poor circumstances for homeowners to trade up, down or across. Someone wishing to trade up or down properties hopes to sell at the highest possible value and also buy at the lowest possible value. This is difficult to achieve in a fast-growing and volatile market. Instead, the best conditions for the property market are where there are slow, predictable price changes.

### About the Kenny Report (1973)

The Report of the Committee on the Price of Building land (known as the Kenny Report) examined possible methods for controlling the price of development land and ensuring that all or a substantial part of the income in the value of land from zoning would be secured for the benefit of the community.

The report observed large increases in the prices paid for serviced land suitable for building and for potential building land near cities and towns in the state. The high price of land was noted as one of the causes of high house prices then. The report observed that from 1963 to 1971 average price of serviced land in County Dublin increased by 530 percent while the Consumer Price Index increased by 64 percent over the same period.

The report recommended the acquisition of newly zoned land by local authorities at 1.25 times its previous value and to resell this to builders to reduce the price of building land. This model exists in Canada and other European countries (O'Mahony Pike Architects, 2021), such as Germany (Ryan-Collins et al., 2017). It also discussed the merits of a site value tax but did not recommend doing so because their remit only referred to building land, consisting of a very small number of sites.

The Kenny Report remains largely unimplemented (Farrell, 2016). However, since there is a surplus of zoned land, the current challenge is mostly not about reducing the value uplift from re-zoning land from agriculture to commercial or residential use. Instead, the current challenge mainly relates to the price of and the underuse of sites that are already zoned for residential and commercial purposes. See An Taisce (2012) and O'Mahony (2018) on over-zoning.

### The limits of stamp duty

After commercial rates, stamp duty is the most important source of property tax in Ireland. For residential property, it is levied at one percent on the first million euros and two percent for the price in excess of one million. The rate is 7.5 percent for non-residential property.

It may be appropriate to reduce stamp duty on property and increase site value tax to compensate. Pike (2006) proposes reducing stamp duty to a level which fulfils its original objective of providing funds for the proper registration of a property. He proposes introducing site value tax and using the proceeds to reduce stamp duty, remove capital gains tax and reduce commercial rates.

Site value tax is a better tax than stamp duty for several reasons. Stamp duty is a more volatile revenue source since is reliant on the number of transactions occurring, the current sale value of property and the type of property currently being sold. Site value tax would also broaden the tax base by applying it to all property owners rather than just those people who are purchasing property (Monaghan, 2010).

Stamp duty also discourages property transactions, in the same way that a financial transaction tax (also called a Tobin tax) is intended to discourage high frequency trading (Becchetti, 2014). Stamp duty also discourages renewal of towns and cities. It penalises conversion relative to new build residential and discourages people from moving closer to their place of work. Stamp duty also discourages downsizing for older people (Pike, 2006).

Muellbauer (2005) describes stamp duty as a 'bad tax' because it discourages transactions where both parties would benefit. Part of the current housing shortage is that people are in the wrong homes. Many young families live in houses that are too small and many older people in houses that are too large. A relatively large and stable number of transactions are good for society as they allow better matching between households and housing requirements. Similarly, stamp duty makes it more expensive to combine multiple sites to make large developments.